\documentclass{article}
\usepackage{scalefnt}
\usepackage{hyperref}
\usepackage{spconf,amsmath,graphicx}
\usepackage{hyperref, url}
\usepackage{pgf-pie}
 \usepackage{booktabs}
 \usepackage{multirow}
 \usepackage{caption}
\usepackage{subcaption}
 \usepackage[english]{babel}
 \usepackage[utf8]{inputenc}

\title{A Two-stage U-Net for high-fidelity denoising of historical recordings}
\name{Eloi Moliner and Vesa V\"alim\"aki\sthanks{This research is part of the activities of the Nordic Sound and Music Computing Network---NordicSMC (NordForsk project no.~86892).}}
\address{Acoustics Lab, Dept.~Signal Processing and Acoustics, Aalto University, FI-02150 Espoo, Finland\\ eloi.moliner@aalto.fi\\}
\begin{document}
%
\maketitle
\begin{abstract}

Enhancing the sound quality of historical music recordings is a long-standing problem. This paper presents a novel denoising method based on a fully-convolutional deep neural network. A two-stage U-Net model architecture is designed to model and suppress the degradations with high fidelity. The method processes the time-frequency representation of audio, and is trained using realistic noisy data to jointly remove hiss, clicks, thumps, and other common additive disturbances from old analog discs. The proposed model outperforms previous methods in both objective and subjective metrics. The results of a formal blind listening test show that real gramophone recordings denoised with this method have significantly better quality than the baseline methods. This study shows the importance of realistic training data and the power of deep learning in audio restoration.

\end{abstract}
%
\section{Introduction}
\label{sec:intro}
Audio recording technology has evolved dramatically since its invention in the 19th century \cite{godsill_digital_1998, Esquef2008}. In comparison to current audio files, due to technological constraints, early recordings have poor quality and are affected by several kinds of degradation, such as the characteristic hiss and clicks \cite{godsill_digital_1998, Esquef2008}. While this material can be digitized, suppressing the disturbances in the recordings is not trivial. Audio restoration has been extensively studied during the last few decades \cite{godsill_digital_1998}. 

This paper deals with the denoising of historical recordings using a novel deep neural network. The approach taken here suppresses colored noises and rumble as well as impulsive events. The denoising problem has been previously tackled with various techniques, such as Wiener filtering, wavelets \cite{Berger1994}, and spectral substraction \cite{boll1979suppression,Ephraim1985Speech}, which only affected stationary noise. The removal of clicks and thumps had to be treated independently by firstly detecting them and then interpolating the missing samples \cite{godsill_digital_1998, rund_evaluation_2021, Carvalho2021}. 
With the rise of machine learning, deep neural networks have been successfully applied for different kinds of audio restoration goals, such as speech enhancement  \cite{isik_poconet_2020, defossez_real_2020}, bandwidth extension \cite{kuleshov_audio_2017}, audio inpainting \cite{marafioti_gacela_2020}, and low-bitrate audio restoration \cite{deng_exploiting_2020}.  

This work is inspired by Li et al.~\cite{li_learning_2020}, who proposed a neural network model for historical music denoising based on a time-frequency U-Net architecture.  Their approach not only relied on a supervised reconstruction loss, but also incorporated a set of adversarial discriminators. Although their experiments showed promising results, a subjective evaluation did not show a performance gain using the adversarial losses. This paper focuses on the fully-supervised approach, setting aside the adversarial strategy. Our main contributions are twofold: we propose a refined model architecture based on two U-Net stages, and we use more consistent and realistic training data than the previous work, which help the model to generalize to real music recordings.





This paper is structured as follows. Sec.~2 discusses the training data, Sec.~3 introduces the two-stage U-Net architecture, Sec.~4 presents the results of objective evaluations and a formal blind listening test, and Sec.~5 concludes.

\section{Clean and Noisy Data Collection}

We train our model using simulated data, with artificially added noise, as is usual in supervised approaches. The input noisy segments $X$ are generated by mixing the clean examples $Y$ with noise $N$: 
\begin{equation}
X=\beta(Y+ \alpha N),    
\end{equation}
where $Y$ and $N$ come from two separate datasets described later on in this section, $\alpha$ is a signal-to-noise-ratio (SNR) scaling factor, and $\beta$ is a level scaling factor. Both $\alpha$ and $\beta$  are useful for data augmentation and to strengthen the model performance in different noise and level conditions. We set the SNR and $\beta$ as log-uniform random variables between [2\,dB, 20\,dB] and [$-$6\,dB, 4\,dB], respectively. The clean examples $Y$ (scaled with the same $\beta$), on the other hand, are treated as targets, to which the reconstruction loss is calculated.


 
 \subsection{Dataset of Clean Classical Music}
 We have chosen to use classical music recordings as training examples, because there is a large quantity of royalty-free high-quality classical-music recordings available on the internet, and that this way we avoid adding anachronisms to our simulated data, as classical-music pieces have remained unchanged over time. We use a refined and extended version of the MusicNet dataset \cite{thickstun_learning_2017}. The oldest recordings with poor audio quality have been discarded, as they contained background noises that could bias the training. In addition to the solo piano and string ensemble recordings from MusicNet, we added further orchestral and opera recordings \cite{archive}. The resulting dataset contains about 40\,h of  classical music.



\subsection{Gramophone Record Noise Dataset}

A key factor for the success of the denoising method is the realism of the noise data. With this idea in mind, we built a dataset of noise segments extracted from the ``The Great 78 Project" \cite{78project}, a large collection of publicly available digitized 78 RPM (rounds per minute) gramophone records \cite{archive}. The noise excerpts we included in the dataset consist of a mixture of degradations from different sources: electrical circuit noise such as hiss, ambient noise from the recording environment, low-frequency rumble noise caused by the turntable, and noises caused by the irregularities in the storage medium in the form of clicks and thumps.

%

Automatically extracting the noise-only segments from digitized recordings in low-SNR conditions is not a trivial task. Our attempts to implement a classifier based on energy thresholds, as in \cite{li_learning_2020}, resulted in outliers, such as very soft music passages, reverberation tails, fading at the beginning and end of musical pieces, or other recorded sounds that could be misclassified as noise. Such outliers harm the training efficiency and the testing veracity, giving inaccurate results. 

With the interest of implementing a robust noise segment extractor, we use a neural network-based binary classifier with a similar architecure as in \cite{sehgal_convolutional_2018}, which we train with a smaller fraction of manually extracted only-noise segments. The resulting dataset contains 139 min of noises, divided into 2430 segments from 1386 different recordings. The recordings used to generate the dataset are dated between the years 1902 and 1966.  We refer the reader to the companion webpage\footnote{http://research.spa.aalto.fi/publications/papers/icassp22-denoising/} for further details on the dataset and the neural-network-based noise selection.
The input length of the denoising model was set to 5\,s, and the shorter noise samples had to be extended. Considering that the minimum segment length of 2\,s is wide enough compared with the receptive field of the denoiser model, there was no advantage in applying a randomized concatenative synthesis or a more refined texture synthesis approach. Instead, we simply repeated the noise samples along time using a fixed crossfade time of 0.5\,s between consecutive noise sequences. This way we not only saved computing time, but also kept the slowly varying periodic patterns defined by the revolution period, which is 0.77\,s in 78-RPM recordings.
 


\section{Two-Stage Music Denoising model}

\begin{figure}[t]
    \centering
    \includegraphics[scale=1]{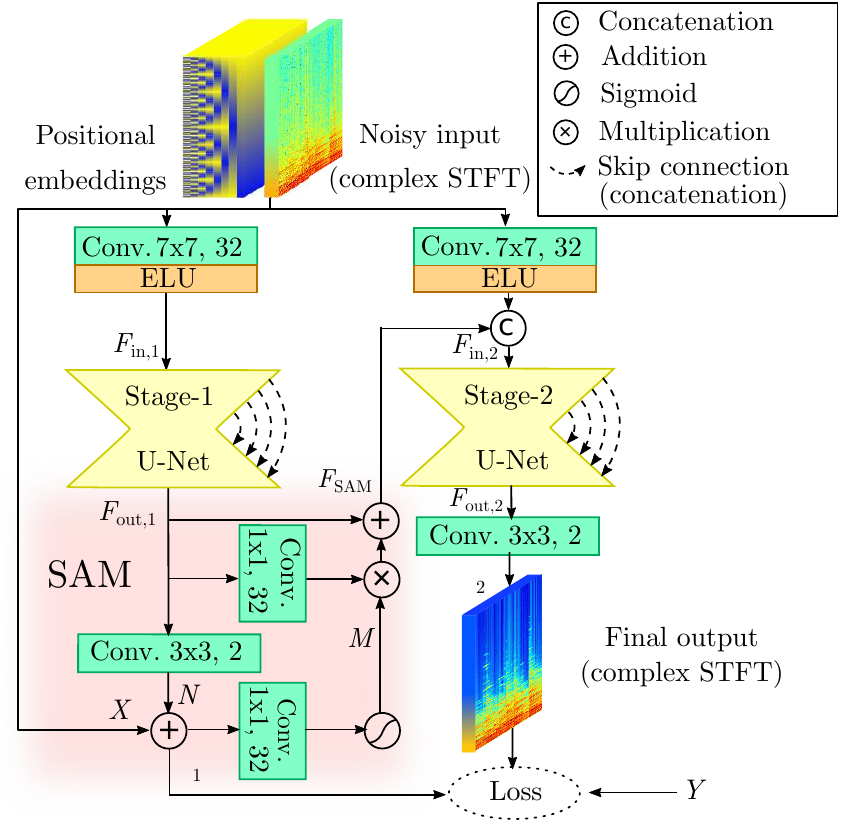}
    \caption{Two-stage denoising model with the SAM module.}
    \label{denoising_framework}
\end{figure}

This section presents a spectrogram-based fully convolutional model for the denoising of historical music recordings. Inspired by recent advances in image restoration \cite{zamir_multi-stage_2021}, we apply a two-stage U-Net architecture with a supervised attention module (SAM), as shown in Fig.~\ref{denoising_framework}. The two-stage approach decomposes the denoising task into two phases. Both stages contain the same U-Net subnetwork, but their inputs and training objectives differ. The goal is to train the first stage to model the residual noise, whereas the second one is trained to effectively denoise and refine the noisy input using the output features from the first stage, thus minimizing the appearance of musical noise artifacts.

The input audio signal, sampled at 44.1\,kHz, is processed in a short-time Fourier transform (STFT) representation, treating its real and imaginary parts as separate real-valued channels. A Hamming window of 2048 samples and a hop size of 512 samples are used in the STFT.  To provide the network frequency information in the first layers, we append frequency-positional embeddings \cite{isik_poconet_2020} to the input data as 10 extra channels. In each stage, the resulting 12-channel input is first processed through an early-feature extractor, consisting of a convolutional layer followed by an exponential linear unit (ELU) \cite{clevert_fast_2016} non-linearity, as shown in Fig.~\ref{denoising_framework}. In the first stage, the extracted features $F_{\text{in},1}$ are directly fed to the U-Net subnetwork. In the second stage, the input features $F_{\text{in},2}$ are generated by concatenating a set of additional features  $F_{\text{SAM}}$, coming from the SAM module in the previous stage.

The purpose of SAM \cite{zamir_multi-stage_2021} is to make the network propagate only the most relevant features to the second stage, discarding the less useful ones. The estimated residual-noise signal $\hat{N}$ is generated from the U-Net output features $F_{\text{out},1}$ by means of a 3$\times$3 convolutional layer. The output of the first stage $\hat{Y}_1$ is then calculated as $\hat{Y}_1=X+\hat{N}$, where $X$ is the input spectrogram. The attention-guided features $F_{\text{SAM}}$ are computed following the scheme in Fig.~\ref{denoising_framework} and using the attention masks $M$, which are directly calculated from $\hat{Y}_1$ with a 1x1 convolution and a sigmoid function. The final denoised output $\hat{Y}_2$ is generated by processing the output features of the second U-Net $F_{\text{out},2}$ with a 3$\times$3 convolutional layer.

We apply ground-truth supervision at the output of each stage by minimizing the mean absolute error of both outputs. The reconstruction loss function is then defined as
\begin{equation}
   \mathcal{L} =\frac{1}{K} \sum_k \left(|\hat{Y}_1^k-Y^k|+|\hat{Y}_2^k-Y^k|\right),
\end{equation}
where $Y$ is the clean spectrogram and $K$ is the total number of STFT bins. The entire model is trained for 300\,000 steps with a batch size of 8. We use the Adam optimizer \cite{kingma_adam_2017} with the parameters $\beta_1=0.5$, $\beta_2=0.9$, and a variable learning rate starting with $1\times10^{-4}$ and dividing it by a factor of 10 after every 100\,000 steps. Normalization strategies are not used, as neither batch normalization nor weight normalization provided any performance gain in our experiments.

\begin{figure}[t]
\centering
 \begin{subfigure}[b]{0.61\columnwidth}
         \centering
         \includegraphics{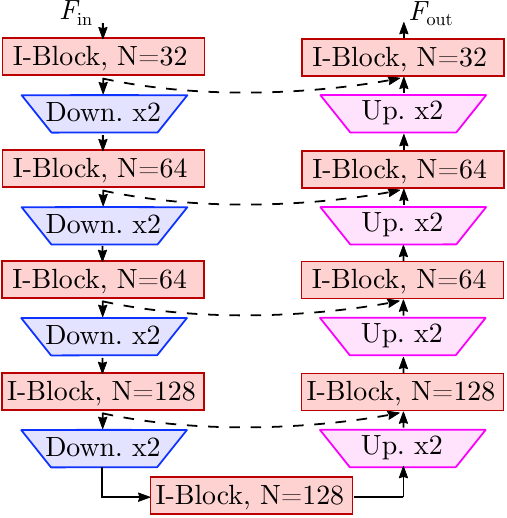}
         \caption{}
         \label{unet}
     \end{subfigure}
     \begin{subfigure}[b]{0.36\columnwidth}
         \centering
         \includegraphics{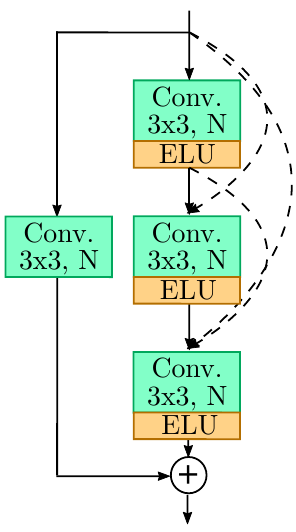}
         \caption{}
         \label{intermediate}
     \end{subfigure}
     \caption{(a) U-Net subnetwork and (b) I-Block.}
\end{figure}

\subsection{U-Net Subnetworks}
The U-Net is widely used for computer vision tasks \cite{Ronneberger2015U-net, zamir_multi-stage_2021}, as well as for audio processing \cite{kuleshov_audio_2017, isik_poconet_2020, defossez_real_2020, li_learning_2020, Choi2020InvestigatingUW}. The U-Net subnetworks used in this work have a symmetric encoder-decoder structure with four downsamplers and upsamplers, as shown in Fig.~\ref{unet}. In each scale, we apply an intermediate block (I-Block), detailed in Fig.~\ref{intermediate}, which  consists of a three-layer DenseNet block \cite{huang_densely_2018} with a residual connection. The outputs at each encoder I-Block are connected to the input of its respective I-Block on the decoder side via concatenative skip connections. We also add an I-Block at the output of the fourth downsampler. The downsampling layers are strided convolutions with a stride of 2$\times$2, kernel size 4$\times$4, and the same number of filters as their subsequent I-Block (see Fig.~\ref{unet}). The upsampling layers in the decoder are transposed convolutions with the same hyperparameters as their symmetric downsamplers. Although the use of transposed convolutions is sometimes discouraged due to the checkerboard artifacts they may produce \cite{Odena}, our experiments showed that these artifacts were undetectable at the later stages of the training.

\section{Experiments and Results}

The quality of the results is evaluated using both objective and subjective methods. The proposed two-stage U-Net denoiser is compared with a traditional denoising method and the state-of-the-art deep-learning method of Li et al.~\cite{li_learning_2020} (STFT-SEANet). The traditional method used is the Ephraim and Malah log-spectral amplitude estimator \cite{Ephraim1985Speech} in combination with autoregressive model-based click detection and least-squares autoregressive interpolation \cite{godsill_digital_1998} (LSA-C). 

\begin{table}[]
\caption{Average objective difference results, where higher is better. The best result in each column in highlighted.}
\label{averages}
\begin{tabular}{@{}ll|lcc@{}}
\toprule
\textbf{SNR}                 & \textbf{Method} & $\Delta$SNR    & $\Delta$PEAQ  & $\Delta$PEMO-Q \\ \midrule
\multirow{3}{*}{3\,dB}  & LSA-C           & 8.97           & 0.08          & 0.23           \\
                             & STFT-SEANet     & 15.28          & 0.60          & 0.86           \\
                             & Proposed        & \textbf{15.89} & \textbf{0.92} & \textbf{1.00}  \\ \midrule
\multirow{3}{*}{10\,dB} & LSA-C           & 7.68           & 0.24          & 0.37           \\
                             & STFT-SEANet     & 13.51          & 0.89          & 1.14           \\
                             & Proposed        & \textbf{14.13} & \textbf{1.16} & \textbf{1.27}  \\ \bottomrule
\end{tabular}%
\end{table}
\begin{figure*}[t]
         \centering
         \includegraphics[trim= 7 7.5 8 7, clip, width=0.9\textwidth]{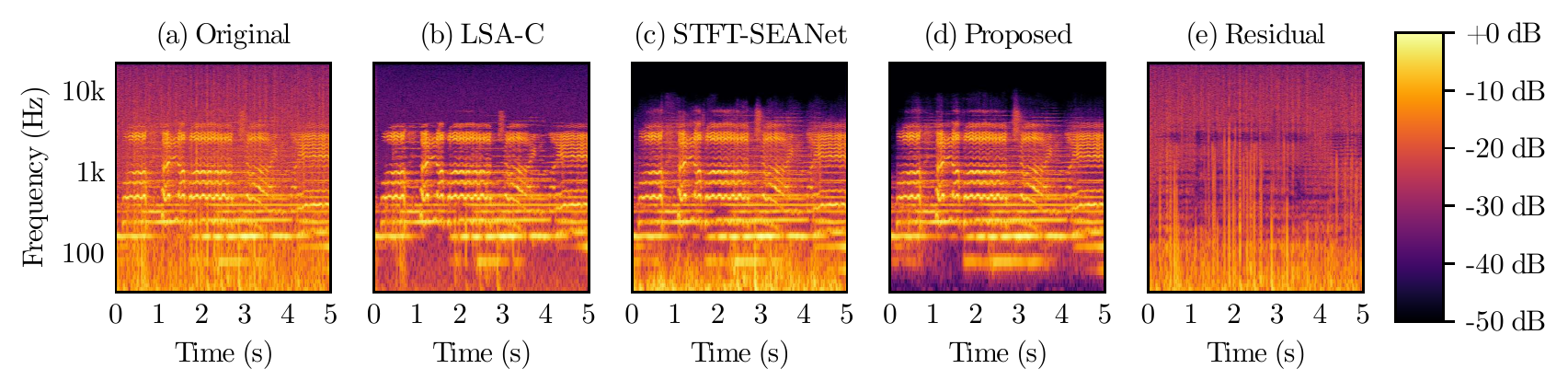}
         \caption{Log-spectrograms  of (a) one of the listening test examples, (b,c,d) denoised versions, and (e) the residual noise.}
         \label{logspectrograms}
\end{figure*}
\subsection{Objective Evaluation}\label{objective}
We evaluate the proposed method with a test set consisting of a total of 66.6 min of 5-s audio mixtures generated with artificially added noise. We balance the test set to equally represent the four main subgenres appearing in the clean classical-music dataset: Solo piano, string ensemble, orchestral, and opera. We evaluate the test set with very adverse (SNR $\sim$ 3 dB) and moderate noise conditions (SNR $\sim$ 10 dB). We report the average gain between the noisy input and the denoised result for three different objective metrics, SNR ($\Delta$SNR) and two perceptual metrics, $\Delta$PEAQ \cite{ITURPEAQ} and $\Delta$PEMO-Q \cite{PEMOQ}.

For this evaluation, an attempt was made to replicate STFT-SEANet as well as possible when using the same training conditions and data, as for the proposed method. The average results of the comparison of the proposed method, STFT-SEANet, and LSA-C are summarized in Table~\ref{averages}. As can be seen, the proposed method outperforms the others in all the evaluated metrics, confirming the performance gain of our two-stage U-Net model against STFT-SEANet's model architecture. The objective results also show a clear difference between both deep-learning-based methods and LSA-C.

\subsection{Subjective Evaluation}

We conducted a blind listening test to evaluate the performance of the proposed method in real recordings. The test method was a multiple-stimuli test in which, unlike the MUSHRA recommendation \cite{ITURmushra}, the original noisy recording was the reference signal, since a clean reference was unavailable. The participants were asked to rate on a scale from 0 to 100 the quality of the denoising in four different conditions. One of the conditions was the hidden reference, which in this case was expected to be rated as 0 (no denoising). The three remaining conditions were denoised versions of the reference using the same methods as in Sec.~\ref{objective}: the proposed method, STFT-SEANet, and LSA-C. 

We used six audio examples of 5\,s each, which were published by the authors of STFT-SEANet \cite{li_learning_2020}. We evaluate their results using reconstruction loss only, as the authors reported this to be their best version. Thus, in contrast with the objective evaluation, we do not evaluate our implementation of the STFT-SEANet, but the actual results obtained by Li et al.~\cite{li_learning_2020} using their own training methodology. 

The LSA-C method requires a noise-only segment to estimate the noise spectral density, which, unfortunately, was unavailable for most examples. However, as can be seen in Fig.~\hyperref[logspectrograms]{3e}, the proposed method isolates the noise fairly by just computing the residual between the denoised output of the two-stage model and the original noisy signal. These examples were shown to the LSA-C method. 

Fig.~\hyperref[logspectrograms]{3} shows a log-spectrogram representation of one of the test examples and its different denoised versions. As can be noticed, the original signal  (Fig.~\hyperref[logspectrograms]{3a}) is widely affected by noise. LSA-C significantly reduces the noise level, but it leaves a noise floor in the spectrum (Fig.~\hyperref[logspectrograms]{3b}). STFT-SEANet achieves a better performance in the higher frequency bands, but it fails to attenuate the low-frequency rumble (Fig.~\hyperref[logspectrograms]{3c}). Finally, the proposed method successfully suppresses the different kinds of noises in the entire frequency range (Fig.~\hyperref[logspectrograms]{3d}).

\begin{figure}[t]
    \centering
    \includegraphics{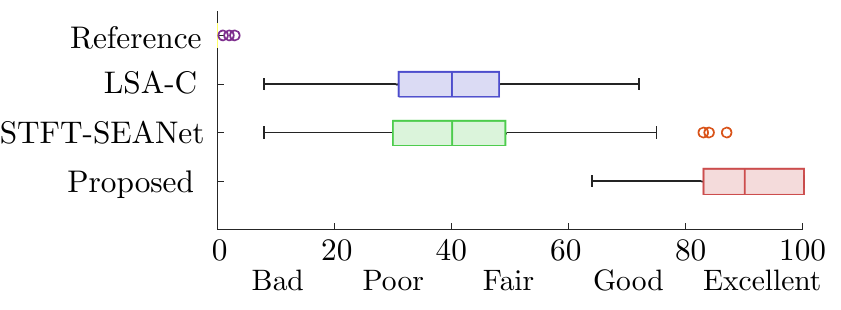}
    \caption{Box-plot visualization of the listening test results.}
    \label{resultssubj}
\end{figure}

Altogether, 13 subjects, with an average age of 29, participated in the listening test. All of them had normal hearing and experience in listening tests. The average results are represented in Fig.~\ref{resultssubj}. The listeners rated the proposed method as ``excellent'' (median 90) with a significant difference compared to the others. None of the subjects had trouble detecting the hidden reference. Some participants found the low-frequency rumble disturbing in the STFT-SEANet condition, in some examples even more than the steady noise of LSA-C. This explains the disappointing results of STFT-SEANet, which obtained a median score of 40, which is midway between ``poor'' and ``fair'', the same as LSA-C. The test scores, the audio files used, and further examples are available\footnotemark[\value{footnote}].

Considering that the rumble noises did not appear in our implementation of the STFT-SEANet, we speculate that they may be caused by a bias in the Li et al. \cite{li_learning_2020} training data. Thus, we attribute the large score differences to a better data collection procedure and the use of the realistic Gramophone Record Noise Dataset.

\section{Conclusions}
This paper presented a high-fidelity audio denoiser for historical recordings based on a deep neural network. The method consists of a two-stage U-Net model that processes audio in a time-frequency representation. The model was trained in a fully-supervised manner using artificially-added noises extracted from real gramophone recordings. The proposed method outperforms previous methods, and in a blind listening test, real gramophone recordings denoised with it were evaluated as significantly better than the compared baselines.

\bibliographystyle{IEEEbib}
\bibliography{refsICASSP, refs}

\end{document}